\input{aipcheck}

\documentclass[
    ,final            
  ]
  {aipproc}

\layoutstyle{8x11double}

\usepackage{unites2e}

\begin{document}

\title{VERITAS Observations of the Arrival Directions of the Highest Energy Cosmic Rays}


\author{J. Holder}{
  address={Dept. of Physics and Astronomy and the Bartol Research Institute, University of Delaware, DE 19716, USA}
}

\author{for the VERITAS Collaboration}{
  address={see http://veritas.sao.arizona.edu/ for a list of collaborators}
}

\begin{abstract}

The recent discovery by the Pierre Auger collaboration of anisotropy
in the arrival directions of the highest energy cosmic rays,
correlated with the positions of nearby active galactic nuclei \cite{Auger07, longAuger07},
encourages the search for counterpart TeV gamma-ray
emission. Approximately half of the sky viewed by the southern
hemisphere Pierre Auger experiment is also visible at reasonable
elevations for the northern hemisphere gamma-ray telescope array,
VERITAS. We report on first observations by VERITAS of regions
associated with the arrival directions of ultra-high energy cosmic ray
events.

\end{abstract}

\maketitle


\section{Introduction}
\begin{figure}
  \includegraphics[height=.3\textheight]{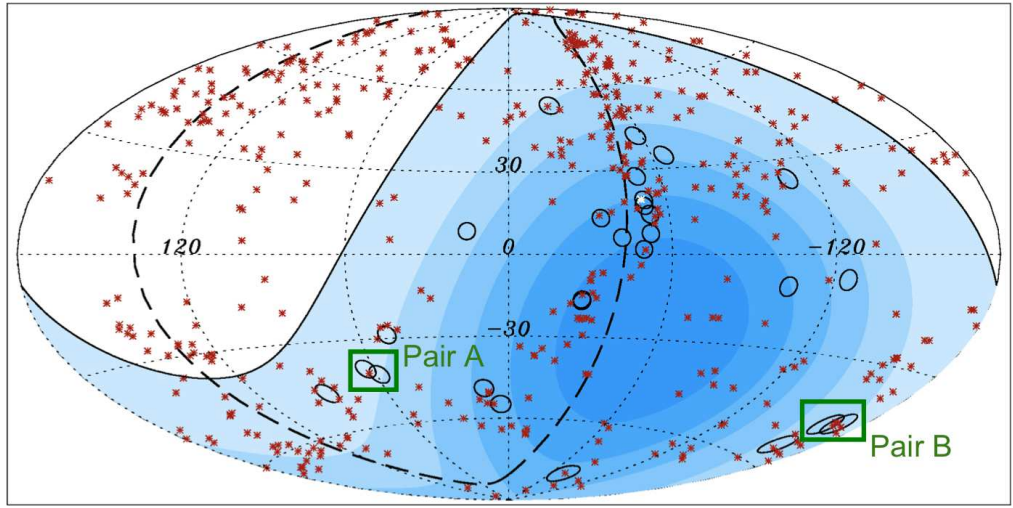}
  \caption{ \label{auger_map} The Pierre Auger Observatory sky map
  (modified from $^1$), showing the celestial sphere in galactic
  coordinates with circles of radius $3.1^{\circ}$ centered at the
  arrival directions of the 27 cosmic rays with highest energy
  detected by the Pierre Auger Observatory \cite{Auger07}. The positions of the 472
  AGN (318 in the field of view of the Observatory) with redshift
  $z\leq0.018 (D < 75\U{Mpc})$ from the 12th edition of the catalog of
  quasars and active nuclei \cite{Veron06} are indicated by red
  asterisks. The solid line draws the border of the field of view
  (zenith angles smaller than $60^{\circ}$). Darker color indicates
  larger relative exposure. Each coloured band has equal integrated
  exposure. The dashed line is, for reference, the super-galactic
  plane. Centaurus A, one of our closest AGN, is marked in white. The
  two pairs of events of interest for the VERITAS observations are
  marked and labeled.}
\end{figure}

The principle of using gamma rays as ``tracers'' for the sources of
cosmic ray acceleration was one of the original motivations behind the
development of gamma-ray astronomy. If cosmic rays are accelerated to
EeV energies in discrete sources, an enhanced flux of lower energy
photons might also be observed from the direction of these
sources. This could result either from interactions of the ultra-high
energy cosmic ray (UHECR) particles with intergalactic matter and
photon fields, in which case the source flux and size depend strongly
on the unknown strength and structure of the intergalactic magnetic
fields, or from lower energy particle acceleration and gamma-ray
emission processes driven by the same central engine.  The Pierre
Auger Observatory sky map (modified
from\footnote{http:://www.auger.org/news/PRagn/AGN\_correlation.html})
is reproduced in Figure~\ref{auger_map}, consisting of 27 events with
energies above 57 EeV. Sources from the Veron- Cetty and Veron AGN
catalogue \cite{Veron06} with z<0.018 are shown by red stars, the
Auger points are shown by black circles with a radius
$\Psi=3.1^{\circ}$.


\section{VERITAS Observations}

VERITAS \cite{Holder06} is an array of four, $12\U{m}$ diameter imaging atmospheric
Cherenkov telescopes located in Tucson, Arizona
($31^{\circ}40'30''\U{N}, 110^{\circ}57'07''\U{W}$, $1268\U{m}$ above
sea level). Of the 27 UHECR events, 7 have declinations $>-10^{\circ}$,
and so are visible to VERITAS at telescope elevations greater than
$50^{\circ}$. Four arrive in two groups of two (labelled Pair A and
Pair B in Figure~\ref{auger_map}). The properties of the UHECR events
in these pairs are summarized in Table~\ref{events}.

\begin{table}[t]
\begin{tabular}{|c|c|c|c|c|c|c|}
\hline
Pair & Year & Day of Year & Auger zenith angle & E(EeV) & RA & Dec\\
\hline
A &  2005 & 63 & 54.5 & 71  & $331.2^{\circ}$ & $-1.2^{\circ}$\\ 
A &  2007 & 51 & 39.2 & 58  & $331.7^{\circ}$ & $2.9^{\circ}$\\ 
\hline
B &  2006 & 35 & 30.8 & 85  & $53.6^{\circ}$ & $-7.8^{\circ}$\\ 
B &  2006 & 296 & 54.0 & 69 & $52.8^{\circ}$ & $-4.5^{\circ}$\\ 
\hline
\end{tabular}
   \caption
   { 
   \label{events}
Properties of the four events used to select the VERITAS search
regions (see \cite{longAuger07} for a detailed description of the
energy and arrival direction reconstruction).
   }
\end{table}

VERITAS observed these regions of the sky in autumn,
2007. Table~\ref{agn_props} lists AGN from the Veron-Cetty and Veron
catalogue with z<0.05 which were within the field of view during these
observations. The exposure time for Pair A was 200 minutes, for Pair
B, 608 minutes. Observations consist of 20 minute runs in the standard
``wobble'' mode, wherein the putative source is offset by
$0.5^{\circ}$ sequentially to the North, South, East and West from the
centre of the field of view. The mean source elevation angle for these
observations was $52^{\circ}$. Pair A observations were centered on
the position of Q~2207+0122, Pair B observations targeted right
ascension $3^h31^m56.46^s$, declination $-05^{\circ}18'59.1''$: a
point equidistant from NGC~1358 and SDSS~J03302-0532 and approximately
$0.5^{\circ}$ from each.

\begin{table}[t]
\begin{tabular}{|c|c|c|c|c|}
\hline
Field & Source Name  & Alternative Name & $z$ & Classification \\
\hline
Pair A & Q~2207+0122      & PC~2207+0122            & 0.013 & Emission Line         \\
Pair A & Q~2207+0121B     & 2MASX J22102668+0136432 & 0.047 & Emission Line  \\
Pair A & Q~2205+0120      & 2MASX J22080139+0135290 & 0.045 & Emission Line  \\
Pair A & SDSS~J22064+0106 & 2MASX J22062439+0106455 & 0.049 & Seyfert 2    \\
Pair A & Q~2212+0215 & 2MASX J22151024+0230415 & 0.041 & Emission Line   \\
Pair A & Q~2213+0218 & PC~2213+0218 & 0.041 & Emission Line \\
\hline
Pair B &NGC~1358         & -            & 0.013 & Seyfert 2              \\
Pair B &SDSS~J03302-0532 & NGC~1346     & 0.014 & Seyfert 1              \\
Pair B &SDSS~J03349-0548 & 2MASX J03345798-0548536 & 0.018 & Seyfert 1 \\ 
\hline
\end{tabular}
   \caption
   { 
   \label{agn_props}
Properties of AGN from the V\'eron-Cetty \& V\'eron catalogue with $z\leq0.05$ within the field of view of the VERITAS observations. The classification is as given by Simbad, angular size is from NED.
   }
\end{table}

\section{Results}

\begin{figure}
   \resizebox{\textwidth}{!}{%
   \begin{tabular}{cc}
   \includegraphics[height=6.0cm]{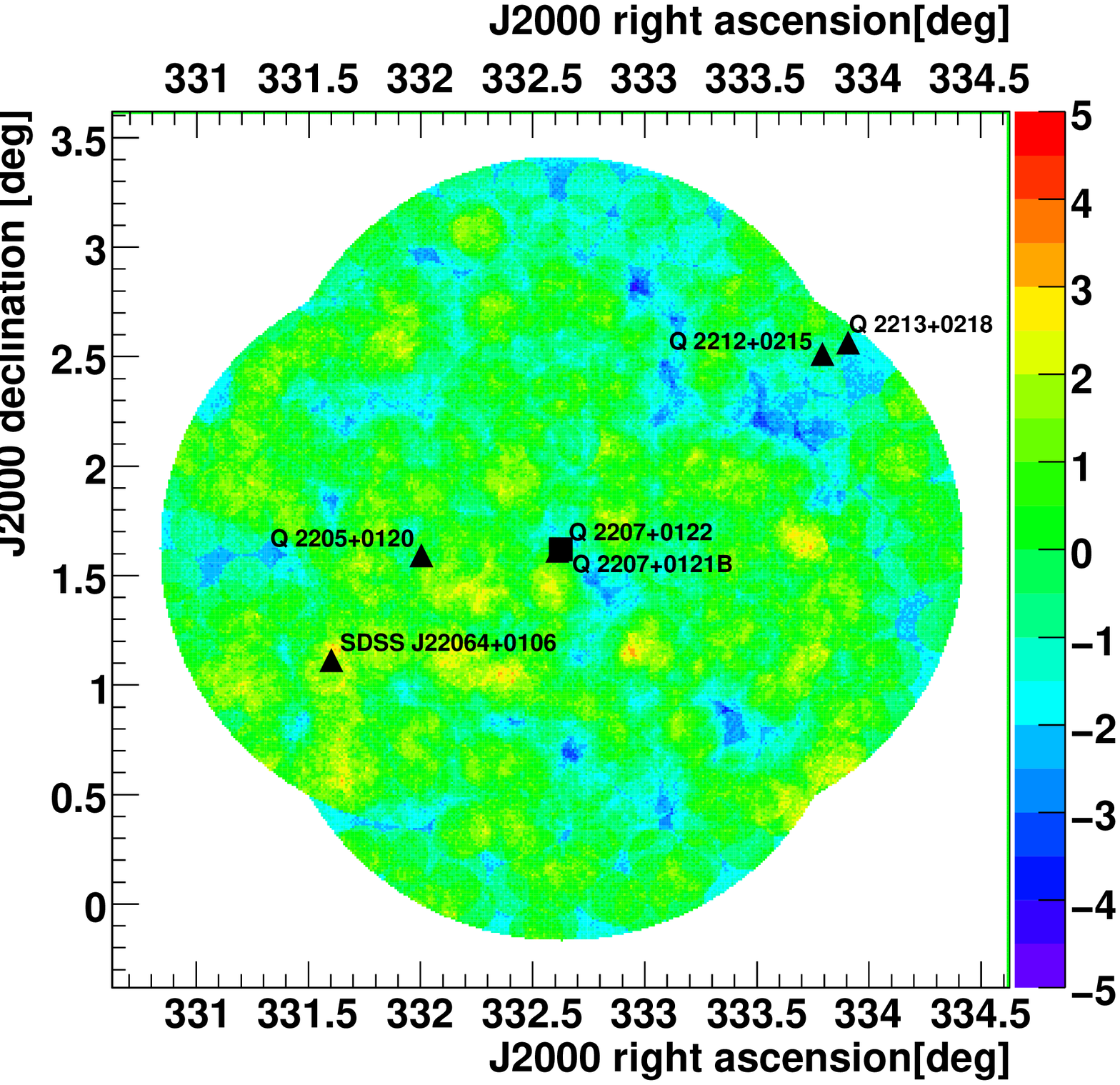}
   \includegraphics[height=6.0cm]{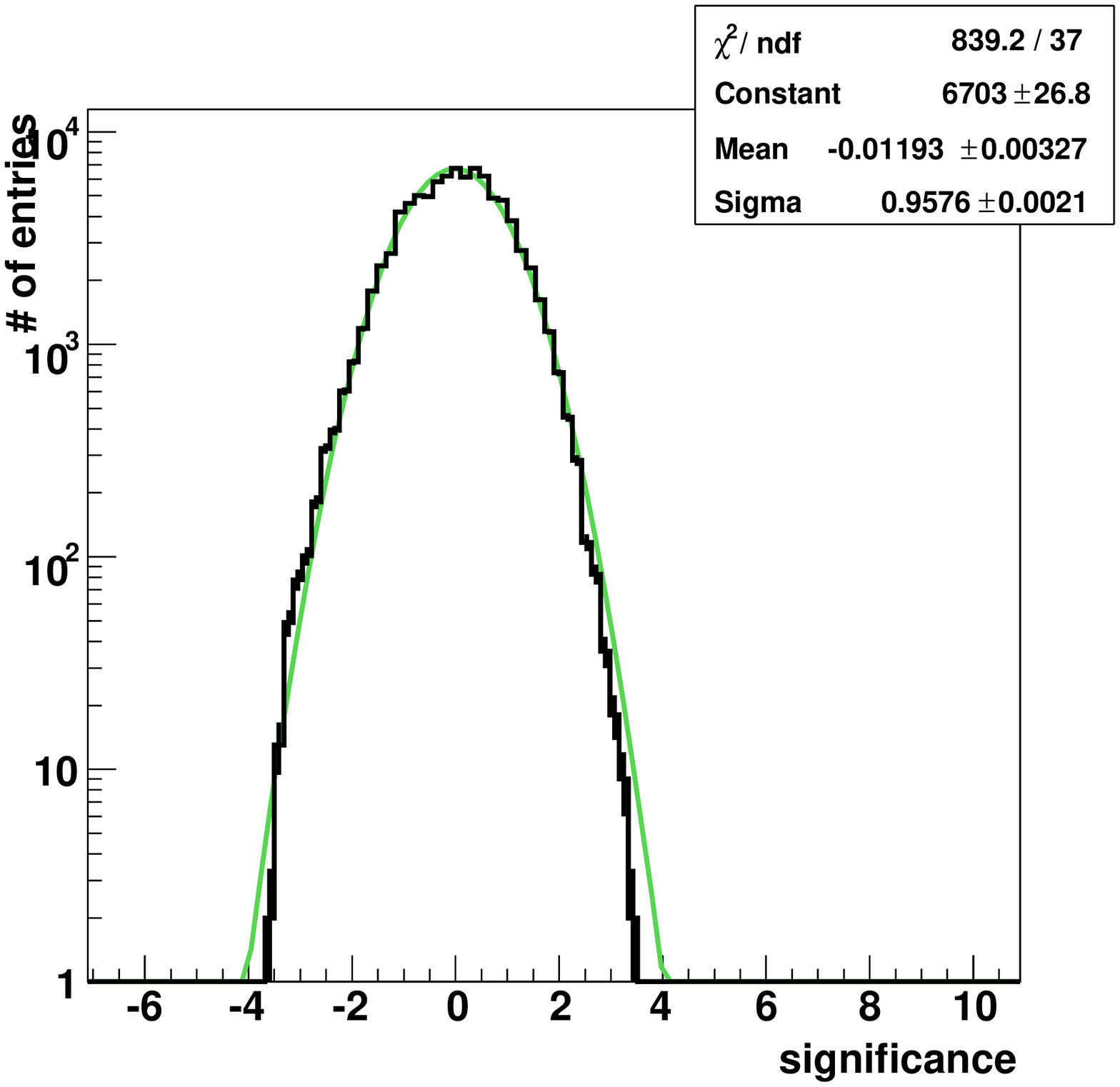}
   \end{tabular}}
   \caption
   { 
   \label{pc2207_plots}
   Significance map in the region of Pair A. AGN at
$z\leq0.018$ are indicated by square markers. More distant
AGN out to $z\leq0.05$ are indicated by triangles. PC~2207+0122 is at the centre of the field; observations were
made with the camera centre offset by $0.5^{\circ}$ from this
position in four different directions. 
   }
   \end{figure}

\begin{figure}
   \resizebox{\textwidth}{!}{%
   \begin{tabular}{cc}
   \includegraphics[height=6.0cm]{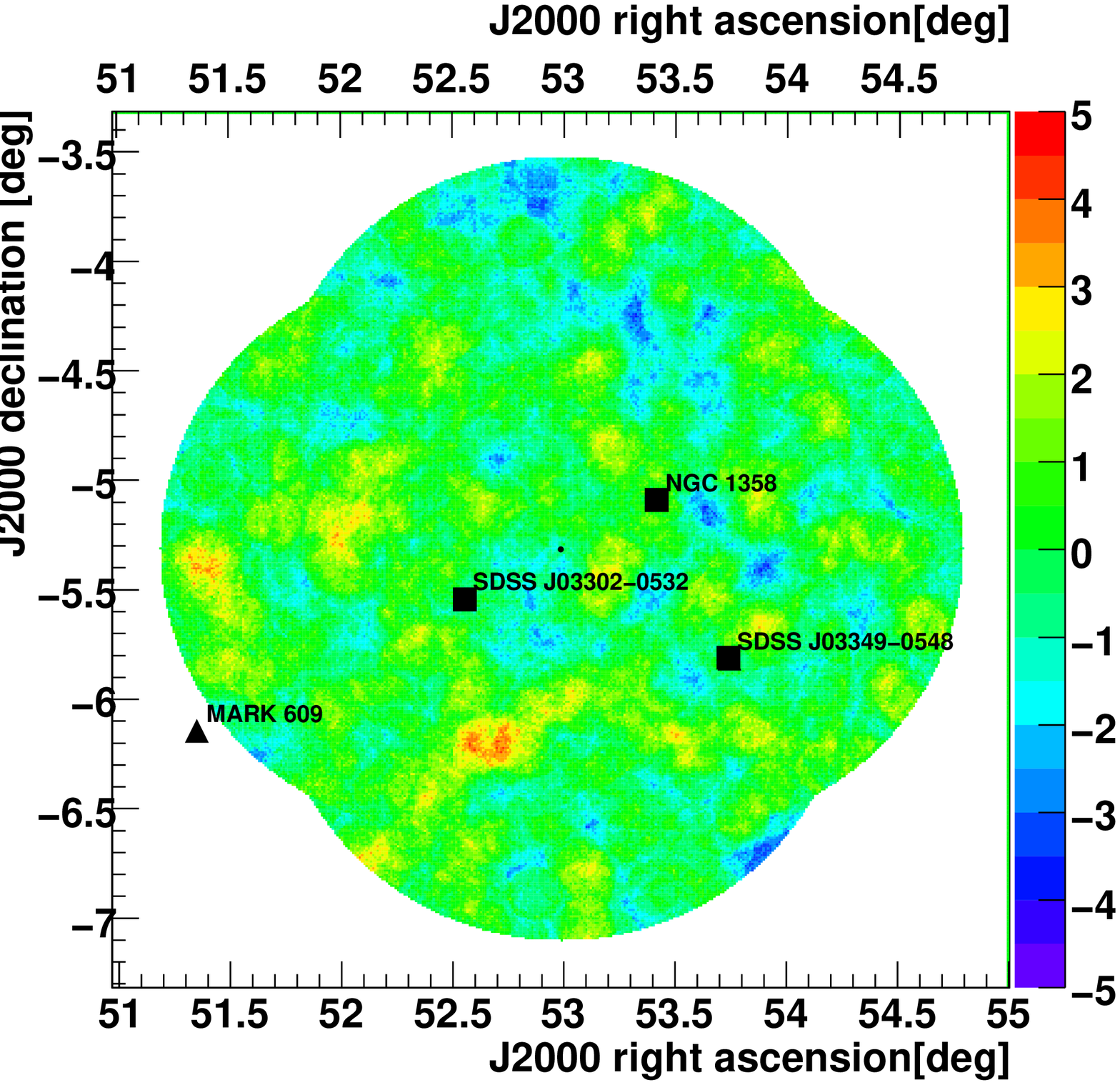}
   \includegraphics[height=6.0cm]{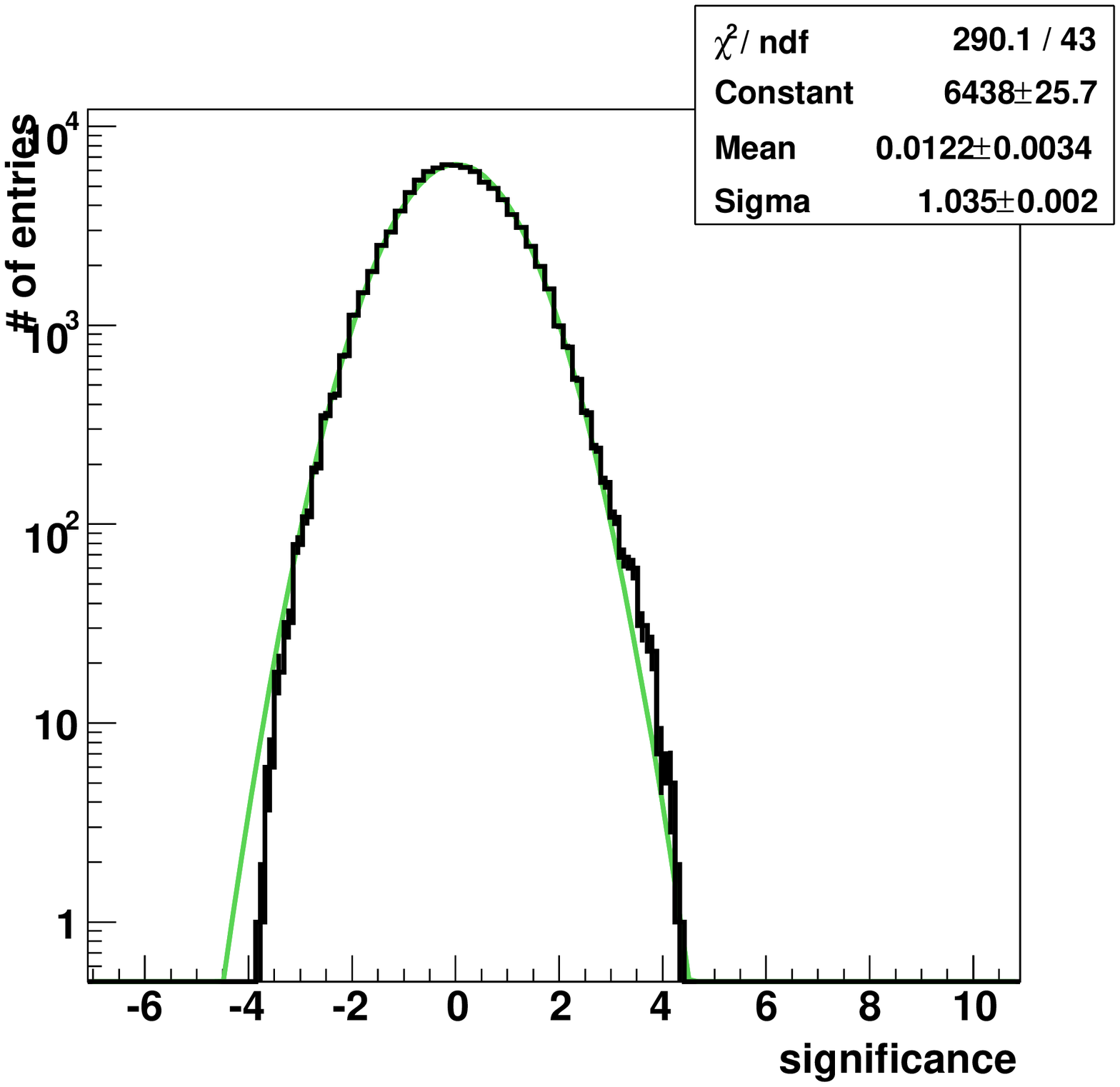}
   \end{tabular}}
   \caption
   { 
   \label{ddt1_plots}
   Significance map in the region of Pair B. AGN at
$z\leq0.018$ are indicated by black square markers. More distant
AGN out to $z\leq0.05$ are indicated by triangles. The
black circle indicates the centre of the field; observations were made with the
camera centre offset by $0.5^{\circ}$ from this position in four
different directions. 
   }
   \end{figure}

The VERITAS observations were analysed using standard analysis tools
(\cite{Acciari08}) and gamma-ray selection cuts optimized for point
sources with a flux of 1\% of the Crab Nebula and a Crab-like
spectrum. Figures~\ref{pc2207_plots} and
\ref{ddt1_plots} show the significance skymaps (left), and the
distribution of significances over all bins in each map (right). No
significant evidence for gamma-ray emission at any position in the
field of view was found.

Table~\ref{agn_results} shows the 99\% confidence integral upper flux
limits for point source emission at each of the AGN positions in the
fields, calculated according to the method of Feldman \& Cousins
\cite{Feldman98}.The upper limits are calculated for an assumed
power-law energy spectrum with a differential spectral index of
$\alpha=-2.5$, and a minimum energy of $500\U{GeV}$, close to the
energy threshold of VERITAS for observations at this angle to the
zenith and for the analysis cuts used.

\begin{table}[t]
\begin{tabular}{|c|c|c|c|c|c|c|}
\hline
 Source Name  & \textit{ON-source} & \textit{OFF-source} & Background    &\multicolumn{2}{|c|}{99\% confidence upper limits} \\
              & (events)           &  (events)           & Normalization & (events) & ($\U{ph}\UU{m}{-2}\UU{s}{-1}>500\U{GeV}$)  \\
\hline
Q~2207+0122       &  4  & 78 & 0.10 & 4.6 & $8.6\times10^{-9}$  \\
Q~2207+0121B      &  4  & 78 & 0.10 & 4.6 & $9.6\times10^{-9}$  \\
Q~2205+0120       &  7  & 73 & 0.10 & 9.5 & $2.3\times10^{-8}$  \\
SDSS~J22064+0106  &  7  & 34 & 0.10 & 13.4 & $4.9\times10^{-8}$  \\
Q~2212+0215       &  3  & 35 & 0.10 & 7.0 & $3.9\times10^{-8}$  \\
Q~2213+0218       &  0  & 10 & 0.10 & 3.8 & $3.8\times10^{-8}$  \\
\hline
\hline
NGC~1358          &  26  & 179 & 0.10 & 19.1 & $1.4\times10^{-8}$  \\
SDSS~J03302-0532  &  13  & 170 & 0.10 & 6.1 & $4.9\times10^{-9}$  \\
SDSS~J03349-0548  &  13  & 137 & 0.10 & 10.8 & $8.3\times10^{-9}$  \\
\hline
\end{tabular}
   \caption
   { 
   \label{agn_results}
Results of VERITAS observations of AGN from the V\'eron-Cetty \& V\'eron catalogue with $z\leq0.05$. Upper limits are calculated for point source emission using the method of Feldman and Cousins \cite{Feldman98}. The upper six sources are associated with Pair A (exposure=200 minutes), the lower three with Pair B (exposure=608 minutes).
}
\end{table}

\section{Conclusions}
We have searched for evidence of gamma-ray emission from two regions
on the sky coincident with the arrival directions of pairs of
ultra-high energy cosmic ray events detected by the Auger
observatory. No significant evidence for emission has been found
within a field of view of $\sim1.7^{\circ}$ radius, and point source
upper limits are given for close AGN in these fields.

Numerous mechanisms exist for the production of a GeV-TeV gamma-ray
flux associated with the production and propagation of ultra-high
energy cosmic rays (e.g. \cite{Aharonian94, Ferrigno05,
Gabici07}). The lack of a detection by VERITAS is not surprising,
however. The flux may be below the sensitivity of these observations,
spatially very extended, located at a large angle to the cosmic ray
arrival directions, or possibly time variable. Nevertheless, the
search for a gamma-ray source population associated with the highest
energy cosmic ray events is worthwhile, as the excellent angular
resolution ($\sim0.01^{\circ}$) of TeV instruments may allow for the
unambiguous identification of the sites of acceleration of these
particles. Ongoing observations by Auger will, in the future, provide
clearer targets for gamma-ray source searches.


\begin{theacknowledgments}
This research is supported by grants from the U.S. Department of
Energy, the U.S. National Science Foundation, the Smithsonian
Institution, by NSERC in Canada, PPARC in the U.K. and by Science
Foundation, Ireland.

This research has made use of the SIMBAD database, operated at CDS,
Strasbourg, France, and the NASA/IPAC Extragalactic Database (NED)
which is operated by the Jet Propulsion Laboratory, California
Institute of Technology, under contract with the National Aeronautics
and Space Administration.

\end{theacknowledgments}



\bibliographystyle{aipproc}   


\end{document}